\documentclass
[superscriptaddress,secnumarabic,amssymb,amsmath,nobibnotes,aps,prd,showkeys,showpacs,nofootinbib,onecolumn,notitlepage,12pt]{revtex4}%
\usepackage{setspace}
\usepackage{color}
\usepackage{amsmath}
\usepackage{amsfonts}
\usepackage{verbatim}
\usepackage{amssymb}
\usepackage{graphicx}
\usepackage[caption=false]{subfig}%
\providecommand{\U}[1]{\protect\rule{.1in}{.1in}}

\newcommand{\be}{\begin{equation}}
\newcommand{\ee}{\end{equation}}

\newcommand{\mincir}{\raise
-3.truept\hbox{\rlap{\hbox{$\sim$}}\raise4.truept\hbox{$<$}\ }}
\newcommand{\magcir}{\raise
-3.truept\hbox{\rlap{\hbox{$\sim$}}\raise4.truept\hbox{$>$}\ }}

\ifx\pdfoutput\relax\let\pdfoutput=\undefined\fi
\newcount\msipdfoutput
\ifx\pdfoutput\undefined\else
\ifcase\pdfoutput\else
\ifx\paperwidth\undefined\else
\ifdim\paperheight=0pt\relax\else\pdfpageheight\paperheight\fi
\ifdim\paperwidth=0pt\relax\else\pdfpagewidth\paperwidth\fi
\fi\fi\fi
\begin{document}
\title{Nonzero Spatial Curvature in Symmetric Teleparallel Cosmology}
\author{Andronikos Paliathanasis}
\email{anpaliat@phys.uoa.gr}
\affiliation{Institute of Systems Science, Durban University of Technology, Durban 4000,
South Africa}
\affiliation{Departamento de Matem\'{a}ticas, Universidad Cat\'{o}lica del Norte, Avda.
Angamos 0610, Casilla 1280 Antofagasta, Chile}

\begin{abstract}
We consider the symmetric teleparallel $f\left(  Q\right)  $-gravity in
Friedmann--Lema\^{\i}tre--Robertson--Walker cosmology with nonzero spatial
curvature. For a nonlinear $f\left(  Q\right)  $ model there exist always the
limit of General\ Relativity with or without the cosmological constant term.
The de Sitter solution is always provided by the theory and for the specific
models $f_{A}\left(  Q\right)  \simeq Q^{\frac{\alpha}{\alpha-1}}%
~,~f_{B}\left(  Q\right)  \simeq Q+f_{1}Q^{\frac{\alpha}{\alpha-1}}$ and
$f_{C}\left(  Q\right)  \simeq Q+f_{1}Q\ln Q$ it was found to be the unique
attractor. Consequently small deviations from STGR can solve the flatness
problem and lead to a de Sitter expansion without introduce a cosmological
constant term. This result is different from that given by the power-law
theories for the other two scalar of the trinity of General Relativity. What
makes the nonlinear symmetric teleparallel theory to stand out are the new
degree of freedom provided by the connection defined in the non-coincidence
frame which describes the nonzero spatial curvature.

\end{abstract}
\keywords{Cosmology; Curvature; $f\left(  Q\right)  $-gravity; Symmetric teleparallel gravity}\maketitle
\date{\today}

\section{Introduction}

\label{sec1}

Cosmological observations reveals the universe's homogeneity and isotropy on
large scales, described by the Friedmann--Lema\^{\i}tre--Robertson--Walker
(FLRW) geometry.{ The value of \ the spatial curvature for the
three-dimensional hypersurface is a debate and recent studies indicate that it
is not zero \cite{cr1,cr2,cr3,cr4,cr5}.} As a solution to the observational
data, it has been proposed that the universe underwent a rapid expansion known
as inflation \cite{guth}. During the inflationary era, the universe was
dominated by a matter source with negative pressure.

There is a category of gravitational models where the Einstein-Hilbert Action
Integral is modified in an effort to include new geometrodynamical terms in
the field equations and explain cosmic acceleration. In this category of
models, the modification of the Action Integral for the gravitational theory
is a direct consequence for the introduction of geometric invariants. For a
four-dimensional manifold $V^{4}$, with a metric tensor $g_{\mu\nu}$ and a
generic connection $\hat{\Gamma}_{\mu\nu}^{\kappa}$, we can define the Riemann
tensor $R_{\;\lambda\mu\nu}^{\kappa}$,~the torsion tensor~$\mathrm{T}%
_{\;\mu\nu}^{\lambda}$ and the non-metricity tensor $Q_{\lambda\mu\nu}%
~$\cite{eisn}. In General Relativity the connection $\hat{\Gamma}_{\mu\nu
}^{\kappa}$ takes the form of the Levi-Civita connection, where $\mathrm{T}%
_{\;\mu\nu}^{\lambda}$ and $Q_{\lambda\mu\nu}$ vanishes, i.e. $\mathrm{T}%
_{\;\mu\nu}^{\lambda}=0$ and $Q_{\lambda\mu\nu}=0$, while the Ricci scalar is
the fundamental scalar for the gravitational Action Integral. Nevertheless, in
the context of Teleparallel Equivalent of General Relativity (TEGR) it holds
$R_{\;\lambda\mu\nu}^{\kappa}=0$ and $Q_{\lambda\mu\nu}=0,$ and the
gravitational the torsion scalar $T$ constructed by the tensor~$\mathrm{T}%
_{\;\mu\nu}^{\lambda}$ plays the role of the gravitational Lagrangian
function$~$\cite{ein28,Tsamp}. However, in Symmetric Teleparallel General
Relativity (STGR) the non-metricity scalar $Q$ is the fundamental
gravitational scalar; while at the same time $R_{\;\lambda\mu\nu}^{\kappa}=0$
and $\mathrm{T}_{\;\mu\nu}^{\lambda}=0~$\cite{Hohmann}. These three
gravitational theories provide the same gravitational field equations, which
is why the Ricci scalar $R$, the torsion scalar $T$, and the non-metricity
scalar $Q$ are known as the trinity of gravity \cite{tr1}. However, this
equivalence of the three theories is lost when nonlinear components are
included in the gravitational Action Integral.

General Relativity is a second-order theory of gravity; however $f\left(
R\right)  $-theory \cite{Buda} is a higher-order theory; specifically
$f\left(  R\right)  $-theory is fourth-order theory and equivalent with a
special case of the Brans-Dicke gravitational model. The power-law $f\left(
R\right)  =R^{\lambda}$, $\lambda\neq0,1$ gravitational model has been widely
studied in cosmological studies \cite{cl1,cl2,cl3}. Indeed, for $\lambda\neq2$
in a spatially flat FLRW the exact solution of the\ Friedmann's equations is
the scaling solution which describes an perfect fluid with constant equation
of state parameter \cite{anbas1}. Nevertheless, for $\lambda=2$, the exact
solution of the cosmological field equations is the de Sitter universe
\cite{sd3}. In the presence of the spatial curvature, or in the presence of
anisotropies, the power-law $f\left(  R\right)  =R^{\lambda}$~model can solve
the flatness problem but the de Sitter universe is recovered only for
$\lambda=2~$\cite{gl1,gl2}. In teleparallelism and in $f\left(  T\right)
$-gravity, the power-law model $f\left(  T\right)  =T^{\lambda}$ leads only to
scaling cosmological solutions \cite{sd4}, while for nonzero spatially
curvature the exponential expansion is provided only for other $f\left(
T\right)  $ functions \cite{sd2}.

In this study, we are interested in the evolution of cosmological dynamics in
symmetric teleparallel $f(Q)-$theory of gravity for the FLRW universe with
nonzero spatial curvature. $f(Q)$-theory has been widely studied by
cosmologists because it can provide an alternative geometric mechanism for
describing cosmological phenomena \cite{f6}. {For various applications of
symmetric teleparallel theory we refer the reader to
\cite{ff1,ff2,ff3,ff4,ff4a,ff4b,ff5,ff6,ff7,nn5,nn7,ww2,qq2,qq3,ffq1,ffq2,ffq3,ffq4,ffq5}
and references therein}. Recently, \ in \cite{qq4} a detailed analysis for the
evolution of cosmological parameters performed within the framework of
$f(Q)$-gravity for a spatially flat FLRW universe. Similar studies have been
presented before \cite{qq4a,qq5,qq6} in the case of the coincidence gauge for
the connection. The novelty of \cite{qq4} lies in assuming the case of the
non-coincidence connection, where new dynamic degrees of freedom are
introduced in the field equation from the connection. This leads to a
higher-order theory of gravity, as also analyzed recently in \cite{qq6}.
Consequently, the additional degrees of freedom provide a more complex
cosmological evolution compared to that given by the coincidence connection.
This means that various epochs of cosmological history can be recovered. An
analysis to use cosmological observations to understand the effects of the
different connections performed recently in \cite{ds00}. {See also the recent
review \cite{rev10}.}

The purpose of this study is to investigate if the nonlinear $f\left(
Q\right)  $-theory can solve the flatness problem and if there exist a
mechanism to provide a de Sitter expansion without a cosmological constant
term. The rapid expansion of the universe, that is, the inflation, it has been
proposed the flatness problem \cite{guth}. During inflation the universe
expands such that the energy density of the curvature term to be neglected.
With this study and the analysis of the dynamics we investigate if in
$f\left(  Q\right)  $-gravity the existence of curvature in the initial
conditions leads to trajectories which describe spatially flat universes. The
structure of the paper follows.

In Section \ref{sec2} we introduce the cosmological model of our consideration
which that of a FLRW spacetime with nonzero spatial curvature in symmetric
teleparallel $f\left(  Q\right)  $-gravity. We present the modified field
equations for arbitrary $f\left(  Q\right)  $ function where we observe that
dynamical degrees of freedom are introduced by the non-coincidence gauge for
the connection. In Section \ref{sec3} we study the evolution of the physical
parameters by studying the phase-space. We compute the stationary points for a
generic function $f\left(  Q\right)  $. However in order to investigate the
stability properties we specify function $f\left(  Q\right)  $. We found that
there exist always a future attractor which correspond to the spatially flat
de Sitter universe. Furthermore, the limit of General Relativity is always
recovered by the nonlinear function $f\left(  Q\right)  $. Consequently, the
nonlinear $f\left(  Q\right)  $-theory can solve the flatness problem and
provide an inflationary mechanism without to introduce a cosmological constant
term. Finally, in Section \ref{sec4} we draw our conclusions.

\section{$f\left(  Q\right)  $-cosmology with nonzero curvature}

\label{sec2}

In cosmological scales the universe is considered to be isotropic and
homogeneous described by the FLRW geometry. In spherical coordinates the
latter spacetime is described by the line element%
\begin{equation}
ds^{2}=-dt^{2}+a\left(  t\right)  ^{2}\left(  \frac{dr^{2}}{1-kr^{2}}%
+r^{2}\left(  d\theta^{2}+\sin^{2}\theta d\phi^{2}\right)  \right)  .
\end{equation}
Parameter $k$ is the spatial curvature of the three-dimensional hypersurface,
$k=1$ describes a closed universe, $k=0$ a flat universe and $k=-1$ describes
an open universe. Function $a\left(  t\right)  $ is the scalar factor which
corresponds the radius of the universe. The Hubble function is defined as
$H=\frac{\dot{a}}{a}$.

The compatible connections $\Gamma_{\mu\nu}^{\kappa}$ for the FLRW geometry
which inherits the spacetime isometries and they describe flat geometry, that
is, $R_{\;\lambda\mu\nu}^{\kappa}\left(  \Gamma\right)  =0$, they have been
derived before in \cite{Hohmann,Heis2}. For zero spatial curvature, $k=0$, it
has been found that there exist three different families of connections,
however for $k\neq0$, there exist a unique connection with nonzero components
in the spherical coordinates \cite{ds1}%
\begin{equation}%
\begin{split}
&  \Gamma_{\;tt}^{t}=-\frac{k+\dot{\gamma}(t)}{\gamma(t)},\quad\Gamma
_{\;rr}^{t}=\frac{\gamma(t)}{1-kr^{2}}\quad\Gamma_{\;\theta\theta}^{t}%
=\gamma(t)r^{2},\quad\Gamma_{\;\phi\phi}^{t}=\gamma(t)r^{2}\sin^{2}(\theta)\\
&  \Gamma_{\;tr}^{r}=\Gamma_{\;rt}^{r}=\Gamma_{\;t\theta}^{\theta}%
=\Gamma_{\;\theta t}^{\theta}=\Gamma_{\;t\phi}^{\phi}=\Gamma_{\;\phi t}^{\phi
}=-\frac{k}{\gamma(t)},\quad\Gamma_{\;rr}^{r}=\frac{kr}{1-kr^{2}},\quad
\Gamma_{\;\theta\theta}^{r}=-r\left(  1-kr^{2}\right)  ,\\
&  \Gamma_{\;\phi\phi}^{r}=-r\sin^{2}(\theta)\left(  1-kr^{2}\right)
,\Gamma_{\;r\theta}^{\theta}=\Gamma_{\;\theta r}^{\theta}=\Gamma_{\;r\phi
}^{\phi}=\Gamma_{\;\phi r}^{\phi}=\frac{1}{r},\Gamma_{\;\phi\phi}^{\theta
}=-\sin\theta\cos\theta,\Gamma_{\;\theta\phi}^{\phi}=\Gamma_{\;\phi\theta
}^{\phi}=\cot\theta.
\end{split}
\label{conk1}%
\end{equation}

For the latter connection the non-metricity scalar reads \cite{qq3} \
\begin{equation}
Q=-6H^{2}+\frac{3\gamma}{a^{2}}H+\frac{3\dot{\gamma}}{a^{2}}+k\left(  \frac
{6}{a^{2}}+\frac{3}{\gamma}\left(  \frac{\dot{\gamma}}{\gamma}-3H\right)
\right)  . \label{ff.00}%
\end{equation}
while the resulting modified Friedmann's equations in $f\left(  Q\right)  $
gravity are \cite{qq3}
\begin{equation}
3H^{2}f^{\prime}(Q)+\frac{1}{2}\left(  f(Q)-Qf^{\prime}(Q)\right)
-\frac{3\gamma\dot{Q}f^{\prime\prime}(Q)}{2a^{2}}+3k\left(  \frac{f^{\prime
}(Q)}{a^{2}}-\frac{\dot{Q}f^{\prime\prime}(Q)}{2\gamma}\right)  =0,
\label{ff.01}%
\end{equation}%
\begin{equation}
-2\left(  f^{\prime}(Q)H\right)  ^{\cdot}-3H^{2}f^{\prime}(Q)-\frac{1}%
{2}\left(  f(Q)-Qf^{\prime}(Q)\right)  +\frac{\gamma\dot{Q}f^{\prime\prime
}(Q)}{2a^{2}}-k\left(  \frac{f^{\prime}(Q)}{a^{2}}+\frac{3\dot{Q}%
f^{\prime\prime}(Q)}{2\gamma}\right)  =0. \label{ff.02}%
\end{equation}

Furthermore, the connection satisfies the equation of motion \cite{qq3}%
\begin{equation}
\dot{Q}^{2}f^{\prime\prime\prime}(Q)\left(  1+\frac{ka^{2}}{\gamma^{2}%
}\right)  +\left[  \ddot{Q}\left(  1+\frac{ka^{2}}{\gamma^{2}}\right)
+\dot{Q}\left(  \left(  1+\frac{3ka^{2}}{\gamma^{2}}\right)  H+\frac
{2\dot{\gamma}}{\gamma}\right)  \right]  f^{\prime\prime}(Q)=0. \label{ff.03}%
\end{equation}

We remark that connection (\ref{conk1}) depends on an function $\gamma\left(
t\right)  $, and that for nonlinear $f\left(  Q\right)  $ function there is
not any coordinate transformation in which the connection can be written in
the coincidence gauge $\Gamma_{\kappa\lambda}^{\mu}=0~$\cite{ds1}. For the
flat space, $k=0$, there are three different connections but only one is
recovered from (\ref{conk1}) by replacing $k=0.$ For the spatially flat
spacetime the coincidence universe exist, and then modified Friedmann
equations are similar with that of the $f\left(  T\right)  $-gravity. Because
for $k\neq0$ there coincidence gauge does not exist, function $\gamma\left(
t\right)  $ can not neglected, which means that the field equations considered
in \cite{ref1} do not describe $f\left(  Q\right)  $-cosmology with nonzero
spatial curvature.

For a linear function $f\left(  Q\right)  $, that is, $f\left(  Q\right)
=Q-2\Lambda$, the field equations are that of General Relativity, where
$\Lambda$ is the cosmological constant term. In this analysis we assume the
power-law function $f\left(  Q\right)  =Q^{\lambda}$, with $\lambda\neq0,1$.

For the power-law function $f\left(  Q\right)  $, the gravitational field
equations (\ref{ff.01})-(\ref{ff.03}) read%
\begin{equation}
3\lambda H^{2}+\frac{1-\lambda}{2}Q-\frac{3\gamma\dot{Q}\lambda\left(
\lambda-1\right)  }{2a^{2}Q}+3k\left(  \frac{\lambda}{a^{2}}-\frac
{\lambda\left(  \lambda-1\right)  \dot{Q}}{2\gamma Q}\right)  =0,
\label{ff.04}%
\end{equation}%
\begin{equation}
-2\left(  \lambda\dot{H}+\lambda\left(  \lambda-1\right)  H\frac{\dot{Q}}%
{Q}\right)  -3\lambda H^{2}-\frac{1-\lambda}{2}Q+\frac{\gamma\dot{Q}%
\lambda\left(  \lambda-1\right)  }{2a^{2}Q}-k\left(  \frac{\lambda}{a^{2}%
}+\frac{3\lambda\left(  \lambda-1\right)  \dot{Q}}{2\gamma Q}\right)  =0,
\label{ff.05}%
\end{equation}%
\begin{equation}
\left(  \lambda-2\right)  \dot{Q}^{2}\left(  1+\frac{ka^{2}}{\gamma^{2}%
}\right)  +Q\left[  \ddot{Q}\left(  1+\frac{ka^{2}}{\gamma^{2}}\right)
+\dot{Q}\left(  \left(  1+\frac{3ka^{2}}{\gamma^{2}}\right)  H+\frac
{2\dot{\gamma}}{\gamma}\right)  \right]  =0. \label{ff.06}%
\end{equation}
where we have assumed that $Q$ is nonzero.

The modified Friedmann's equations can be written equivalently as follows%
\begin{equation}
3\lambda H^{2}=\rho_{Q}-\rho_{k}+\rho_{\gamma}%
\end{equation}%
\begin{equation}
-2\dot{H}-3H^{2}=p_{Q}+\frac{1}{3}\rho_{k}+p_{\gamma}%
\end{equation}
\qquad where we have defined the effective fluid components%
\begin{equation}
\rho_{Q}=\frac{\lambda-1}{2\lambda}Q~,~\rho_{k}=\frac{3k}{a^{2}}~,~
\end{equation}%
\begin{equation}
\rho_{\gamma}=\frac{3}{2}\left(  \frac{\gamma}{a^{2}}+\frac{k}{\gamma}\right)
\left(  \lambda-1\right)  \left(  \ln Q\right)  ^{\cdot},
\end{equation}%
\[
p_{Q}=2\left(  \lambda-1\right)  H\left(  \ln Q\right)  ^{\cdot}-\frac
{\lambda-1}{2\lambda}Q
\]%
\begin{equation}
p_{\gamma}=\frac{1}{2}\left(  -\frac{\gamma}{a^{2}}+\frac{3k}{\gamma}\right)
\left(  \lambda-1\right)  \left(  \ln Q\right)  ^{\cdot}.
\end{equation}

The effective fluid, characterized by components $\rho_{\gamma}$ and
$p_{\gamma}$, incorporates the dynamical terms associated with the function
$\gamma$, which arises from the non-coincidence connection (\ref{conk1}). In
the asymptotic limit where $\left(  \ln Q\right)  ^{\cdot}\rightarrow0$, we
find that $\rho_{\gamma}=0$ and $p_{\gamma}=0$, while $\rho_{Q}=\rho_{Q_{0}}$
and $p_{Q}=-\rho_{Q_{0}}$. Consequently, the dynamical behavior aligns with
that of General Relativity, including a cosmological constant term, where
$\rho_{Q0}$ plays the role of the cosmological constant.

To gain deeper insights into the cosmological dynamics and evolution of the
cosmological parameters in the presence of curvature for $f\left(  Q\right)
$-gravity, we will conduct a comprehensive analysis of the asymptotic dynamics
for the field equations (\ref{ff.01})-(\ref{ff.03}). This detailed examination
will enable us to determine the general behavior of the physical parameters
within the framework of nonlinear symmetric teleparallel theory, while also
shedding light on the influence of non-zero spatial curvature on the
cosmological dynamics.

\section{Phase-space description}

\label{sec3}

We introduce the scalar field $\psi=f^{\prime}\left(  Q\right)  $, such that
to write the field equations (\ref{ff.01})-(\ref{ff.03}) in the following form%

\begin{equation}
3\psi H^{2}+V\left(  \psi\right)  -\frac{3\gamma}{2a^{2}}\dot{\psi}+3k\left(
\frac{\psi}{a^{2}}-\frac{\dot{\psi}}{2\gamma}\right)  =0, \label{ff.10}%
\end{equation}%
\begin{equation}
-2\left(  \psi H\right)  ^{\cdot}-3\psi H^{2}-V\left(  \psi\right)
+\frac{\gamma\dot{\psi}}{2a^{2}}-k\left(  \frac{\psi}{a^{2}}+\frac{3\dot{\psi
}}{2\gamma}\right)  =0, \label{ff.11}%
\end{equation}
\begin{equation}
\left(  1+\frac{ka^{2}}{\gamma^{2}}\right)  \ddot{\psi}+\dot{\psi}\left(
\left(  1+\frac{3ka^{2}}{\gamma^{2}}\right)  H+\frac{2\dot{\gamma}}{\gamma
}\right)  =0, \label{ff.12}%
\end{equation}
in which $V\left(  \psi\right)  =\frac{1}{2}\left(  f(Q)-Qf^{\prime
}(Q)\right)  $ plays the role of the scalar field potential, and $Q=-2V\left(
\psi\right)  _{,\psi}$

Therefore the effective energy density and pressure components related to the
nonlinear $f\left(  Q\right)  $ are~$\rho_{Q}^{eff}=\rho_{Q}+\rho_{\gamma}$,
$p_{eff}=p_{Q}+p_{\gamma}$ with%
\begin{equation}
\rho_{Q}=-\frac{V\left(  \psi\right)  }{\psi}~,~p_{Q}=2H\left(  \ln
\psi\right)  ^{\cdot}~,~
\end{equation}%
\begin{equation}
~\rho_{\gamma}=\frac{3}{2}\left(  \frac{\gamma}{a^{2}}+\frac{k}{\gamma
}\right)  \left(  \ln\psi\right)  ^{\cdot}+\frac{V\left(  \psi\right)  }{\psi
}~,
\end{equation}%
\begin{equation}
p_{\gamma}=\frac{1}{2}\left(  -\frac{\gamma}{a^{2}}+\frac{3k}{\gamma}\right)
\left(  \ln\psi\right)  ^{\cdot}~.
\end{equation}

Finally, in the scalar field description, equation (\ref{ff.00}) reads%
\begin{equation}
-6H^{2}+\frac{3\gamma}{a^{2}}H+\frac{3\dot{\gamma}}{a^{2}}+k\left(  \frac
{6}{a^{2}}+\frac{3}{\gamma}\left(  \frac{\dot{\gamma}}{\gamma}-3H\right)
\right)  +2V\left(  \psi\right)  _{,\psi}=0. \label{ff.13}%
\end{equation}

\subsection{Dimensionless variables}

To analyze the dynamics effectively, we incorporate the use of new dependent
variables%
\[
x=\frac{\dot{\psi}}{2\psi H}~,~y=\frac{V\left(  \psi\right)  }{3\psi H^{2}%
}~,~z=\frac{\gamma}{a^{2}H}~,~\Omega_{k}=\frac{k}{a^{2}H^{2}}~,~\mu=\psi
\frac{V_{,\psi}}{V},
\]
and the new independent variable $\tau=\ln a$. \ Parameter $\tau$ is
introduced in order the trajectories to depend on the size of the universe and
not the time-parameter.

Hence, the field equations (\ref{ff.10}), (\ref{ff.11}), (\ref{ff.12}) and
(\ref{ff.13}) are written in the equivalent form of the first-order dynamical
system%
\begin{align}
\frac{2z\left(  \Omega_{k}+z^{2}\right)  ^{2}}{x}\frac{dx}{d\tau}  &
=\Omega_{k}^{2}z\left(  \Omega_{k}-3+3y\right)  +2\Omega_{k}\left(  \Omega
_{k}-7+3y\right)  z^{3}\nonumber\\
&  +8\left(  \Omega_{k}-1+\mu y\right)  z^{4}+\left(  5+\Omega_{k}+3y\right)
z^{5}-x\left(  \Omega_{k}+z^{2}\right)  ^{2}\left(  3\Omega_{k}-z^{2}\right)
, \label{dn.01}%
\end{align}%
\begin{equation}
\frac{dy}{d\tau}=y\left(  3\left(  1+y\right)  +\Omega_{k}+x\left(  2\left(
1+\mu\right)  +\frac{3\Omega_{k}}{z}-z\right)  \right)  , \label{dn.02}%
\end{equation}%
\begin{align}
2\frac{dz}{d\tau}  &  =x\left(  3\Omega_{k}-z\left(  z-4\right)  \right)
\nonumber\\
&  +\frac{z}{\Omega_{k}+z^{2}}\left(  \Omega_{k}\left(  5+\Omega
_{k}+3y\right)  -4\left(  \Omega_{k}-1+\mu y\right)  z+\left(  \Omega
_{k}-3\left(  1-y\right)  \right)  z^{2}\right)  \label{dn.03}%
\end{align}%
\begin{equation}
\frac{d\Omega_{k}}{d\tau}=\Omega_{k}\left(  1+\Omega_{k}+3y+x\left(
4+\frac{3\Omega_{k}}{z}-z\right)  \right)  , \label{dn.04}%
\end{equation}%
\begin{equation}
\frac{d\mu}{d\tau}=2\mu x\left(  1-\mu+\mu\Gamma\left(  \mu\right)  \right)  ,
\label{dn.05}%
\end{equation}
with algebraic constraint
\begin{equation}
1+\Omega_{k}+y-\frac{x}{z}\left(  \Omega_{k}+z^{2}\right)  =0, \label{dn.06}%
\end{equation}
and function $\Gamma\left(  \mu\right)  $ is defined as
\begin{equation}
\Gamma\left(  \mu\right)  =\frac{V_{,\psi\psi}V}{\left(  V_{,\psi}\right)
^{2}}. \label{dn.07}%
\end{equation}

Therefore, the deceleration parameter is expressed as
\begin{equation}
q\left(  x,y,z,\Omega_{k}\right)  =\frac{1}{2}\left(  1+\Omega_{k}+3y+z\left(
4+\frac{3\Omega_{k}}{z}-z\right)  \right)  .
\end{equation}

With the application of the constraint equation (\ref{dn.06}) the dynamical
system lies on a four dimensional manifold. Hence, by substituting
$y=-1-\Omega_{k}+\frac{x}{z}\left(  \Omega_{k}+z^{2}\right)  $ in
(\ref{dn.01}), (\ref{dn.03}), (\ref{dn.04}) and (\ref{dn.05}) we end with the
system%
\begin{align}
\frac{2z\left(  \Omega_{k}+z^{2}\right)  ^{2}}{x}\frac{dx}{d\tau}  &
=x\left(  \Omega_{k}+z^{2}\right)  \left(  3\Omega_{k}^{2}+4\Omega_{k}%
z^{2}+4\mu z^{3}+z^{4}\right) \nonumber\\
&  -z\left(  \Omega_{k}^{2}\left(  3+\Omega_{k}\right)  +2\Omega_{k}\left(
\Omega_{k}+5\right)  z^{2}+4\left(  \mu+1+\left(  \mu-1\right)  \Omega
_{k}\right)  z^{3}+\left(  \Omega_{k}-1\right)  z^{4}\right)  , \label{dn.10}%
\end{align}%
\begin{align}
\frac{dz}{d\tau}  &  =-\frac{z}{\Omega_{k}+z^{2}}\left(  \left(  \Omega
_{k}-1\right)  \Omega_{k}-2\left(  1+\mu+\left(  \mu-1\right)  \Omega
_{k}\right)  z+\left(  3+\Omega_{k}\right)  z^{2}\right) \nonumber\\
&  +x\left(  3\Omega_{k}+z\left(  2\left(  1-\mu\right)  +z\right)  \right)  ,
\label{dn.11}%
\end{align}%
\begin{equation}
\frac{d\Omega_{k}}{d\tau}=2\Omega_{k}\left(  -\left(  1+\Omega_{k}\right)
+2x\left(  2+\frac{3\Omega_{k}}{z}+z\right)  \right)  , \label{dn.12}%
\end{equation}%
\begin{equation}
\frac{d\mu}{d\tau}=2x\mu\left(  1-\mu+\mu\Gamma\left(  \mu\right)  \right)  .
\label{dn.13}%
\end{equation}

There are two special functional forms of the scalar field potential of
special interest. The power-law potential $V\left(  \psi\right)  =V_{0}%
\psi^{\mu_{0}}$, where equation (\ref{dn.13}) becomes $\frac{d\mu}{d\tau}=0$,
that is, $\mu=\mu_{0}$, and the exponential potential $V\left(  \psi\right)
=V_{0}\psi^{\alpha}e^{\frac{1}{2}\psi_{0}\left(  \ln\psi\right)  ^{2}}$, in
which $\mu\left(  1-\mu+\mu\Gamma\left(  \mu\right)  \right)  =\psi_{0}$. For
the power-law potential the dimension of the dynamical system is reduced by
one; while for the second potential function stationary points exist only when
$x=0$.

That is an important observation because for an arbitrary function $f\left(
Q\right)  $, the existence of a stationary point $\mu=\mu_{0}$ it means that
at the asymptotic limit $f\left(  Q\right)  $ function is described by the
power-law function $Q^{\frac{\mu_{0}}{\mu_{0}-1}}$, which provides the
power-law potential $V\left(  \psi\right)  =V_{0}\psi^{\mu_{0}}$.

That is similar to the quintessence scalar field \cite{cop1} where the two
exponential potential is related to the scaling solution, while the power-law
potential attributes the de Sitter limit \cite{cop2}. Thus, any other
potential function consists the scaling solution at the limit in which the
potential is described by the exponential function, or the de Sitter limit
when the potential is described by the power-law function. As an example we
recall the hyperbolic potential \cite{cop3} which has the power-law and the
exponential limits \cite{cop4}, that is, why the hyperbolic potential provides
de Sitter and scaling solutions for the background equations.

\subsection{Asymptotic solutions at the finite regime}

Now, let us proceed to determine the stationary/equilibrium points of the
system (\ref{dn.10})-(\ref{dn.13}). Each stationary point corresponds to an
asymptotic solution for the field equations. Let $P=\left(  x\left(  P\right)
,z\left(  P\right)  ,\Omega_{k}\left(  P\right)  ,\mu\left(  P\right)
\right)  $ be one such stationary point of the dynamical system, and as a
stationary point, the right-hand sides of equations (\ref{dn.10}%
)-(\ref{dn.13}) become zero.

For an arbitrary potential $V\left(  \psi\right)  $, that is, for an arbitrary
potential function $\Gamma\left(  \mu\right)  $, then if $\mu_{0}$, is a real
root of equation $\mu\left(  1-\mu+\mu\Gamma\left(  \mu\right)  \right)  =0,$
then the equilibrium points of the dynamical system (\ref{dn.10}%
)-(\ref{dn.13}) are
\begin{align*}
P_{1}  &  =\left(  -1,z_{1},-z_{1},\mu_{0}\right)  ,~P_{2}=\left(  \frac{1}%
{2},2,0,\mu_{0}\right)  ,\\
P_{3}  &  =\left(  \frac{5}{1-3\mu_{0}},-1-\mu_{0},0,\mu_{0}\right)  \text{
and }P_{4}=\left(  0,\frac{2\left(  1+\mu_{0}\right)  }{3},0,\mu_{0}\right)  .
\end{align*}
In the special case where $\mu_{0}=2$, there exist the additional equilibrium
point%
\[
P_{5}=\left(  -1,-3,\Omega_{k5},2\right)  .
\]

For the asymptotic solutions described at the letter points we derive the
deceleration parameters%
\[
q\left(  P_{1}\right)  =0~,~q\left(  P_{2}\right)  =1~,~q\left(  P_{3}\right)
=\frac{2\left(  2-\mu_{0}\right)  }{1-3\mu_{0}}~,~q\left(  P_{5}\right)
=-1~,
\]
and for $\mu_{0}=2,~$%
\[
q\left(  P_{5}\right)  =0.
\]

The stationary points have distinct characteristics: Stationary point $P_{1}$
represents a family of Milne-like universes with a scale factor $a\left(
t\right)  =a_{0}t$. Point $P_{2}$ corresponds to a spatially flat FLRW
asymptotic solution, where the effective fluid behaves as a stiff fluid. At
stationary point $P_{3}$, we find a spatially flat self-similar universe with
a deceleration parameter $q\left(  P_{3}\right)  =\frac{2\left(  2-\mu
_{0}\right)  }{1-3\mu_{0}}$. This point exists for $\mu_{0}\neq\frac{1}{3}$
and describes an accelerated universe when $\frac{1}{3}<\mu<2$. Point $P_{4}$
holds special significance as it corresponds to the de Sitter solution.
Lastly, stationary point $P_{5}$ describes Milne-like universes with a scale
factor $a\left(  t\right)  =a_{0}t$. Notably, this result aligns with the
findings in \cite{qq3}, which analyzed the existence of self-similar solutions.

In addition to the equilibrium points mentioned above, there always exist
other stationary points in the system.
\[
P_{6}=\left(  0,0,-1,0\right)  ~,~P_{7}=\left(  0,\frac{2\left(  1+\mu\right)
}{3},0,\mu\right)
\]
where in $P_{7}$, $\mu$ is arbitrary parameter such that the algebraic
expression $\mu\left(  1-\mu+\mu\Gamma\left(  \mu\right)  \right)  $ to be
non-singular. Points~$P_{6}$ and the family of points $P_{7}$ are the only
points which exist in the case where $\mu\left(  1-\mu+\mu\Gamma\left(
\mu\right)  \right)  =0$ has no real solutions. We remark that $P_{4}$ is a
special case of point $P_{7}$. These two points, describe the asymptotic limit
of the theory in the context of General Relativity with or without the
cosmological constant term.

We derive the deceleration parameters%
\[
q\left(  P_{6}\right)  =0\text{ and }q\left(  P_{7}\right)
=-1\text{\thinspace}.
\]
Therefore, point $P_{6}$ is the Milne solution of General Relativity, and
$P_{7}$ are family of de Sitter points.

We note that for arbitrary function $f\left(  Q\right)  $, these are the
possible families of asymptotic solutions described by the theory. For a
specific definition for the function $f\left(  Q\right)  $, the existence
conditions are defined and the stability properties can be examined.

The significance of the General Relativity limit lies in the existence of
points $P_{6}$ and $P_{7}$ for any arbitrary function $f\left(  Q\right)  $.
The novelty introduced by the non-coincidence gauge it follows from the the
inclusion of the gauge function $\gamma$ in the connection. It is important to
note that for the non-flat FLRW spacetime, this represents the only possible
connection, as referred to in \cite{ds1}.

\subsection{Stability properties}

Until now we have discussed only the existence conditions and the physical
properties of the stationary points for the field equations. However, in order
to reconstruct the cosmological history the stability properties of the points
should be determined. Hence, we should specify the potential function
$V\left(  \psi\right)  $. \ In order to demonstrate the stability properties
we consider the following cases for the scalar field potential, $V_{A}\left(
\psi\right)  =V_{0}\psi^{\alpha}$, $V_{B}\left(  \psi\right)  =V_{0}\left(
\psi-1\right)  ^{\alpha}$ and $V_{C}=V_{0}e^{\alpha x}$. The corresponding
$f\left(  Q\right)  $-theories related to these potentials are%
\[
f_{A}\left(  Q\right)  \simeq Q^{\frac{\alpha}{\alpha-1}}~,~f_{B}\left(
Q\right)  \simeq Q+f_{1}Q^{\frac{\alpha}{\alpha-1}}\text{ and }f_{C}\left(
Q\right)  \simeq Q+f_{1}Q\ln Q.
\]

\subsubsection{Power law potential $V_{A}\left(  \psi\right)  =V_{0}%
\psi^{\alpha}$}

For the power-law potential we calculate $\mu=\alpha$ and $\Gamma\left(
\mu\right)  =\frac{\alpha-1}{\alpha}$, that is, the dimension of the dynamical
system is reduced by one (\ref{dn.10})-(\ref{dn.13}). Consequently, we study
the stability of the stationary points on the constant surface $\mu=\alpha$.

The eigenvalues of the linearized system (\ref{dn.10})-(\ref{dn.12}) around
the stationary point $P_{2}$ are, $\left\{  -2,2,3+\mu\right\}  $, which means
that $P_{2}$ is a saddle point. For the point $P_{3}$ the three-dimensional
system provides the eigenvalues $\left\{  \frac{4\left(  2-\mu\right)
}{1-3\mu},\frac{2\left(  \mu-2+\sqrt{4\mu\left(  9+4\mu\right)  -11}\right)
}{1-3\mu},\frac{2\left(  \mu-2-\sqrt{4\mu\left(  9+4\mu\right)  -11}\right)
}{3\mu-1}\right\}  $, from where we can infer that $P_{3}$ is always a saddle
point. For the de Sitter point $P_{4}$ we derive the eigenvalues $\left\{
-2,-3,-5\right\}  $, that is, $P_{4}$ is a future attractor. For the limit of
General Relativity, i.e. point $P_{6}$, the eigenvalues are $\left\{
-2,2,2\right\}  $, thus, the Milne solution is always unstable and $P_{6}$ is
a saddle point. Finally, for the family of points $P_{1}$ the eigenvalues are
$\left\{  0,0,2\left(  2-\mu\right)  \right\}  $. Because of the zero
eigenvalues in order to infer about the stability properties of the point we
shall study the trajectories numerically. \newline

\begin{figure}[ptbh]
\centering\includegraphics[width=1\textwidth]{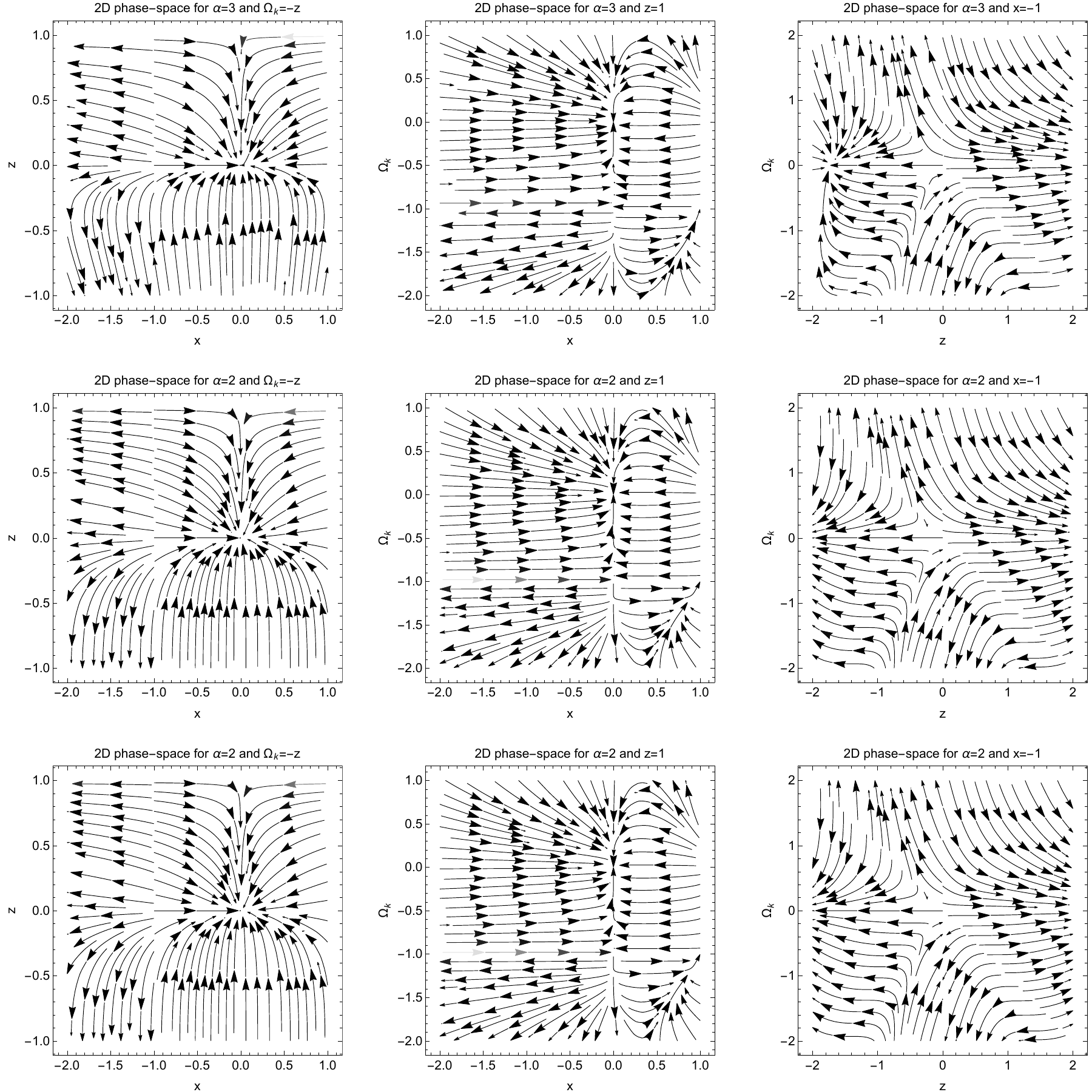}\caption{Two-dimensional
phase-space portraits for the dynaical system (\ref{dn.10})-(\ref{dn.12}) in
the surfaces $\left(  x,z\right)  ,~\left(  x,\Omega_{k}\right)  $ and
$\left(  z,\Omega_{k}\right)  $ where points $P_{1}$ are defined. Figures are
for various values of the free parameter $\alpha$, where we remark that points
$P_{1}$ describe unstable solutions.}%
\label{fig1}%
\end{figure}

In Fig. \ref{fig1} we present two-dimensional phase-portraits for the
dynamical system (\ref{dn.10})-(\ref{dn.12}) for different values of the free
parameters $\alpha$.\ The surfaces are where the family of points $P_{1}$ are
defined. Thus, from the Fig. we observe that $P_{1}$ are always saddle points.

Furthermore, for the special case where $\alpha=2$, that is, $V_{A}\left(
\psi\right)  =V_{0}\psi^{2}$, for the linearized system (\ref{dn.10}%
)-(\ref{dn.13}) around the equilibrium point $P_{5}$ we find the eigenvalues
$\left\{  -1,2,7\right\}  $; we conclude that $P_{5}$ is always a saddle point.

\subsubsection{Power law potential $V_{B}\left(  \psi\right)  =V_{0}\left(
\psi-1\right)  ^{\alpha}$}

For the second potential function $V_{B}\left(  \psi\right)  =V_{0}\left(
\psi-1\right)  ^{\alpha}$ it follows, $\psi=\frac{\mu}{\mu-\alpha}$ and
$\Gamma\left(  \mu\right)  =\frac{\alpha}{\alpha-1}$. Consequently, equation
(\ref{dn.13}) becomes%
\begin{equation}
\frac{d\mu}{d\tau}=\frac{2}{\alpha}x\mu\left(  \mu-\alpha\right)  .
\label{dn1}%
\end{equation}
Indeed, there are two roots of the algebraic equation $\mu\left(  \mu
-\alpha\right)  =0$, they are, $\mu_{1}=0$ and $\mu_{2}=\alpha$.

The dynamical system now exists in four dimensions, and accordingly, there are
four eigenvalues associated with it. In the case of General Relativity, with
or without the cosmological constant, represented by points $P_{6}$ and
$P_{7}$ respectively, the eigenvalues of the linearized system are ${-2, 2, 2,
0}$ for $P_{6}$ and ${-2, -3, -5, 0}$ for $P_{7}$.

As a result, point $P_{6}$ always functions as a saddle point, while our focus
shifts towards the de Sitter solutions at point $P_{7}$. The existence of the
zero eigenvalue, as derived from equation (\ref{dn1}), suggests that the
trajectories lie along the $\mu$-direction. Despite this, the physical
properties of the asymptotic solution, such as the deceleration parameter,
remain unchanged. Although it may appear that $P_{7}$ also behaves as a saddle
point, it is essential to recognize that this family of points defines a space
where the trajectories reside on a three-dimensional surface, specifically
corresponding to the de Sitter universe.

For $\mu_{1}=0$, and points $P_{1}$ we calculate the eigenvalues $\left\{
0,0,2,4\right\}  $, hence $P_{1}$ are saddle points. Moreover, the eigenvalues
related to point $P_{2}$ are ${ -1,-2,2,3} $, while the eigenvalues related to
$P_{3}$ are ${ 8,-10,-4+2\sqrt{11}i,-4+2\sqrt{11}i } $, which means that
$P_{2}$ and $P_{3}$ are saddle points. Finally, for $P_{4}$ it holds the same
discussion with $P_{7}$.

On the surface for $\mu_{1}=\alpha$, with omit the presentation of the
eigenvalues, however we find that the stability properties of the stationary
points $P_{2}$, $P_{3}$ and~$P_{4}$ are the same as before. Specifically
$P_{2}$, $P_{3}$ are saddle points and $P_{4}$ has the same behaviour as
points $P_{7}$. However, for points $P_{1}$ we derive the eigenvalues
$\left\{  0,0,4-2\alpha,-2\right\}  $, where the phase-space portraits
\ref{fig1} are valid, and the same conclusion holds.

\subsubsection{Exponential potential $V_{C}\left(  \psi\right)  =V_{0}%
e^{\alpha\psi}$}

Consider now the exponential potential $V_{C}\left(  \psi\right)
=V_{0}e^{\alpha\psi}$, from where we calculate $\mu=\alpha x$ and
$\Gamma\left(  \mu\right)  =1$. Hence, equation (\ref{dn.13}) reads%
\begin{equation}
\frac{d\mu}{d\tau}=2\mu x. \label{dn2}%
\end{equation}

The stationary points consist of those with $\mu_{0}=0$ and the ones
corresponding to the limit of General Relativity, namely $P_{6}$ and $P_{7}$.
As observed previously, the dynamical system reveals that the de Sitter
universe stands as the only attractor, while all the other asymptotic
solutions are unstable and represented by saddle points.

\section{Poincare variables}

\label{sec3a}

The variables $\left\{  x,z,\Omega_{k},\mu\right\}  $ of the dynamical system
(\ref{dn.10})-(\ref{dn.13}) are not constraint and they can take values in all
the range of real numbers; that is, they can reach infinity. To examine the
existence of stationary points at the infinity we introduce Poincare variables.

For simplicity on our calculations we select $\mu=\alpha$ which corresponds to
the power-law function $f\left(  Q\right)  =Q^{\alpha}$.

\subsection{Closed universe $\Omega_{k}>0$}

For $\Omega_{k}>0$ we define the variables
\[
x=\frac{x}{\sqrt{1-X^{2}-W^{2}-Z^{2}}}~,~z=\frac{Z}{\sqrt{1-X^{2}-W^{2}-Z^{2}%
}}~,~~\Omega_{k}=\frac{W^{2}}{1-X^{2}-W^{2}-Z^{2}},
\]
where now the deceleration parameter reads%
\[
q\left(  X,W,Z\right)  =\frac{Z\left(  Z^{2}+X^{2}-1\right)  +X\left(
3W^{2}+Z\left(  Z+2\sqrt{1-X^{2}-W^{2}-Z^{2}}\right)  \right)  }{Z\left(
1-X^{2}-W^{2}-Z^{2}\right)  },
\]
and the field equations read%
\[
\frac{dX}{dT}=\chi\left(  X,W,Z,\alpha\right)  ~,~\frac{dW}{dT}=\omega\left(
X,W,Z,\alpha\right)  ~,~\frac{dZ}{dT}=\zeta\left(  X,W,Z,\alpha\right)
\]
where $Z\neq0$ and $dT=\sqrt{1-X^{2}-W^{2}-Z^{2}}d\tau.$

At the infinity limit, $1-X^{2}-W^{2}-Z^{2}$, the stationary points
$A=A\left(  X\left(  A\right)  ,W\left(  A\right)  ,Z\left(  A\right)
\right)  $ of the field equations are%
\[
A^{\pm}=\left(  0,0,\pm1\right)  .
\]

The deceleration parameters of the asymptotic solutions at these stationary
points are%
\[
q\left(  A^{\pm}\right)  =0\text{.}%
\]
Consequently, the points $A^{\pm}$ correspond to Milne-like solutions.
Although the eigenvalues of the linearized points are all zero, numerical
solutions indicate that these points are consistently unstable.

It is worth noting that a family of transition points exists on the surface
$Z=0$, comprising the four points $C_{1}^{\pm}=\left(  \pm1,0,0\right)  $ and
$C_{2}^{\pm}=\left(  0,\pm1,0\right)  $. These transition points represent Big
Rip singularities, with points $C_{1}^{-}$ and $C_{2}^{+}$ corresponding to
such scenarios, while points $C_{1}^{+}$ and $C_{2}^{-}$ describe a super
collapse universe.

\begin{figure}[ptbh]
\centering\includegraphics[width=0.9\textwidth]{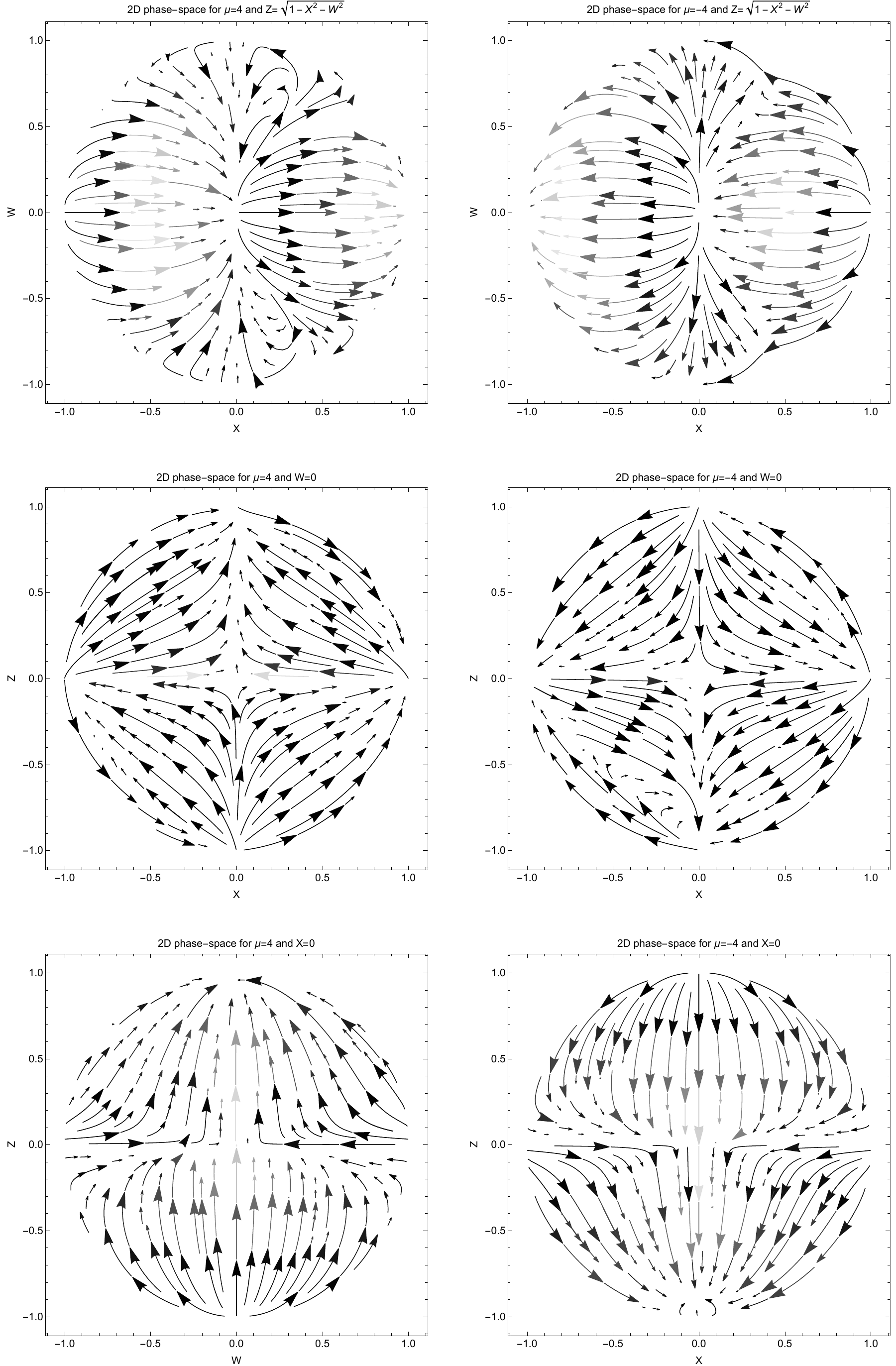}\caption{Two-dimensional
phase-space portraits for the dynaical system (\ref{dn.10})-(\ref{dn.12}) at
the infinity for a closed FLRW spacetime, on diferent surfaces from where it
is clear that there are not attractors at the finite regime. Figures of the
left column are for $\mu=4$, and figures of the right column are for $\mu=-4$.
}%
\label{fig1B}%
\end{figure}

In Fig. \ref{fig1B}, we provide two-dimensional phase-space portraits for the
dynamical system (\ref{dn.10})-(\ref{dn.12}) and the power-law theory at
infinity, observed on different surfaces. It is evident from the figures that
there are no attractors within the finite regime.

\subsection{Open universe $\Omega_{k}<0$}

\begin{figure}[ptbh]
\centering\includegraphics[width=0.9\textwidth]{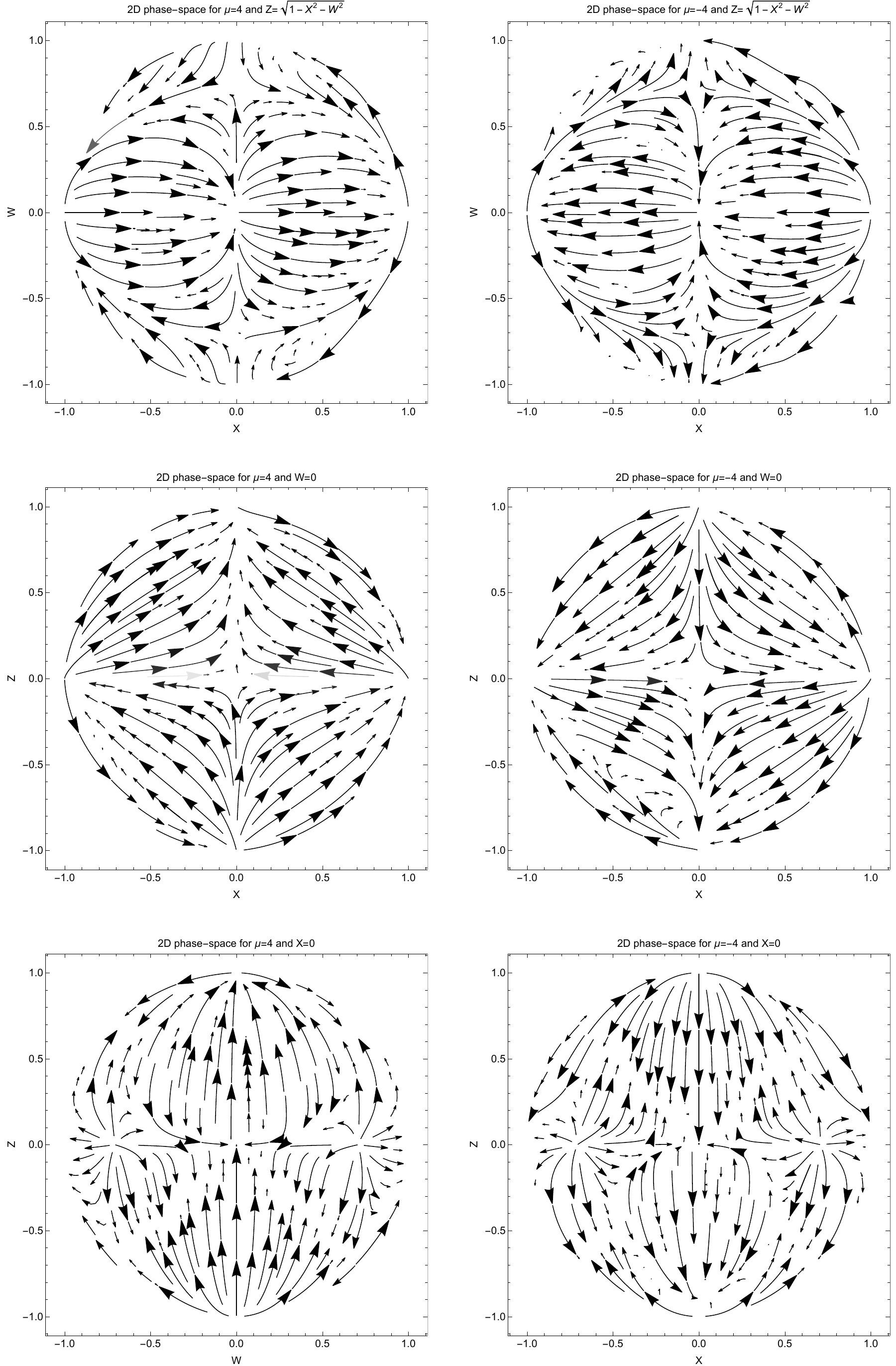}\caption{Two-dimensional
phase-space portraits for the dynaical system (\ref{dn.10})-(\ref{dn.12}) at
the infinity for an open FLRW spacetime, on diferent surfaces from where it is
clear that there are not attractors at the finite regime. Figures of the left
column are for $\mu=4$, and figures of the right column are for $\mu=-4$. }%
\label{fig1C}%
\end{figure}

In the case of an open universe, that is, $\Omega_{k}<0$, the Poincare
variables are
\[
x=\frac{x}{\sqrt{1-X^{2}-W^{2}-Z^{2}}}~,~z=\frac{Z}{\sqrt{1-X^{2}-W^{2}-Z^{2}%
}}~,~~\Omega_{k}=-\frac{W^{2}}{1-X^{2}-W^{2}-Z^{2}}.
\]
We follow the same procedure as before and we find that the there is not any
attractor at the infinity. Milne universes are described by the saddle points
$A_{3}^{\pm}$. In Fig. \ref{fig1C} we present phase-space portraits of the
dynamical system for different values of the free parameters.

It is important to note that the results presented above hold true for a
general $f\left(  Q\right)  $ function. However, the existence of the
stationary points is contingent upon the additional equation (\ref{dn.13}).
While this aspect is of particular interest from a mathematical analysis
perspective, it does not significantly contribute to the physical discussion
regarding the viability of $f\left(  Q\right)  $-theory.

\section{Conclusions}

\label{sec4}

We studied the dynamics of the cosmological field equations in the symmetric
teleparallel $f\left(  Q\right)  \,$-theory, for a FLRW geometry with nonzero
spatial curvature. We study the phase-space and we determined the stationary
points such that to construct all the asymptotic solutions and the
cosmological history.

Through our analysis, we determined the equilibrium points for the general
nonlinear function $f\left(  Q\right)  $, and notably, we found that the
theory always yields the Milne spacetime as a limit, which is consistent with
General Relativity. Additionally, we recovered the de Sitter spacetime,
representing the limit of General Relativity when a cosmological constant term
is present. Furthermore, our investigation yielded novel asymptotic solutions
originating from the geometric terms of the theory. These solutions are
associated with Milne-like scenarios and spatially flat scaling spacetimes.

We examined the stability characteristics of the stationary points for three
distinct functions in our study: $f_{A}\left(  Q\right)  \simeq Q^{\frac
{\alpha}{\alpha-1}}$, $f_{B}\left(  Q\right)  \simeq Q+f_{1}Q^{\frac{\alpha
}{\alpha-1}}$, and $f_{C}\left(  Q\right)  \simeq Q+f_{1}Q\ln Q$. Through our
investigation, we made a significant observation that the spatially flat de
Sitter universe serves as the exclusive attractor within the theoretical framework.

In addition to the above analysis, we delved into the stability properties of
the stationary points at infinity specifically for the power-law function
$f_{A}\left(  Q\right)  $. To achieve this, we utilized Poincare variables in
our study.

Therefore, nonlinear symmetric teleparallel theory can solve the flatness
problem and lead to de Sitter expansion without introduce a cosmological
constant term. This conclusion for the $f\left(  Q\right)  $-gravity is
different from that of the nonlinear theories constructed by the other two
scalar of the trinity of General Relativity, that is, of the $f\left(
R\right)  =R^{\lambda}$ and of the $f\left(  T\right)  =T^{\lambda}$ theories.

In power-law $f\left(  R\right)  $-gravity the de Sitter universe is recovered
when $\lambda=2~$\cite{sd3}, while in the power-law $f\left(  T\right)
$-gravity only scaling cosmological solutions are provided \cite{sd4}.
However, what makes the nonlinear symmetric teleparallel theory to stand out
is the introduction of the additional dynamical variables provided by the
non-coincidence connection and described by the function $\gamma$. While
someone will expect that the power-law theory $f_{A}\left(  Q\right)  \simeq
Q^{\frac{\alpha}{\alpha-1}}$ will not recover General Relativity, as in the
case of spatially flat FLRW in the coincidence gauge, that is not true,
because of the contribution by the non-coincidence gauge terms.

{While this research was still under consideration by the journal, a preprint
with a similar study was made available online \cite{arx1}. In the mentioned
work, the authors adopted the Ansatz }$\gamma\simeq a\left(  t\right)  ${,
restricting their analysis to a specific surface. As a result, their findings
may not be applicable to arbitrary initial conditions or the original problem
in its entirety. Moreover, in their specification of the function }$f\left(
Q\right)  ${, the authors ended up with a system that has more equations than
the original degrees of freedom. Consequently, the analysis presented in
\cite{arx1} is deemed incomplete. In contrast, in our work, we considered a
generic function }$\gamma${, allowing for a more comprehensive exploration of
the topic, and our results remain valid for any set of initial conditions.}

Before we conclude we would like to discuss what will be evolution of the
perturbations in the case of the non-coincidence gauge. Assume the spatially
flat case, $k=0$, then for this connections, the field equations
(\ref{ff.01})-(\ref{ff.02}) becomes%
\begin{equation}
3H^{2}f^{\prime}(Q)+\frac{1}{2}\left(  f(Q)-Qf^{\prime}(Q)\right)
=\rho_{\gamma}, \label{cc.01}%
\end{equation}%
\begin{equation}
-2\left(  f^{\prime}(Q)H\right)  ^{\cdot}-3H^{2}f^{\prime}(Q)-\frac{1}%
{2}\left(  f(Q)-Qf^{\prime}(Q)\right)  =p_{\gamma}. \label{cc.02}%
\end{equation}
with%
\begin{equation}
\rho_{\gamma}=\frac{3\gamma\dot{Q}f^{\prime\prime}(Q)}{2a^{2}}~~,~p_{\gamma
}=-\frac{\gamma\dot{Q}f^{\prime\prime}(Q)}{2a^{2}}~. \label{cc.03}%
\end{equation}
From (\ref{cc.03}) we derive the equation of state parameter $p_{\gamma
}=-\frac{1}{3}\rho_{\gamma}$.

The lhs parts of equations (\ref{cc.01}), (\ref{cc.02}) form the modified
field Friedmann's equations of $f\left(  Q\right)  $-theory in the coincidence
gauge. Therefore the field equations for the non-coincidence gauge introduce
an matter component which is coupled to $\psi=f^{\prime}\left(  Q\right)  $.
Because of that analogue it is expected the evolution of the perturbations to
be similar with that of the coincidence gauge with the extra matter term
\cite{per1,per2}. Where $\gamma\rightarrow0$, that is, on the surface with
$~z=0$, the limit of the coincidence gauge is recovered.

\begin{acknowledgments}
This work was financially supported in part by the National Research
Foundation of South Africa (Grant Numbers 131604). The author thanks the
support of Vicerrector\'{\i}a de Investigaci\'{o}n y Desarrollo
Tecnol\'{o}gico (Vridt) at Universidad Cat\'{o}lica del Norte through
N\'{u}cleo de Investigaci\'{o}n Geometr\'{\i}a Diferencial y Aplicaciones,
Resoluci\'{o}n Vridt No - 098/2022.
\end{acknowledgments}


\begin{thebibliography}{99}                                                                                               %


\bibitem {cr1}E. Di Valentino, A. Melchiorri and J. Silk, Nature Astronomy 4,
196 (2020)

\bibitem {cr2}S. Vagnozzi, E. Di Valentino, S. Gariazzo, A. Melchiorri, O.
Mena and J. Silk, Phys. Dark Univ. 33, 100851 (2021)

\bibitem {cr3}S. Vagnozzi, A. Loeb and M. Moresco, Astrophys. J. 908, 84 (2021)

\bibitem {cr4}S. Dhawan, J. Alsing and S. Vagnozzi, Mon. Not. Roy. Astron.
Soc. 506, L1 (2021)

\bibitem {cr5}A. Glanville, C. Howlett, T. M. Davis, Mon. Not. Roy. Astron.
Soc. 517, 3087 (2022)

\bibitem {guth}A. Guth, Phys. Rev. D 23, 347 (1981)

\bibitem {wald}R.M. Wald, Phys Rev. 28, 2118 (1983)

\bibitem {eisn}L.P. Eisenhart, Non-Riemannian Geometry, Dover Books on
Mathematics, Dover Publications (2012)

\bibitem {ein28}A. Einstein 1928, Sitz. Preuss. Akad. Wiss. p. 217; ibid p. 224

\bibitem {Tsamp}M. Tsamparlis, Phys. Lett. A {75,} 27 (1979)

\bibitem {Hohmann}M. Hohmann, Phys. Rev. D 104, 124077 (2021)

\bibitem {tr1}J. Beltr\'{a}n-Jim\'{e}nez, L. Heisenberg and T.S. Koivisto,
Universe 5, 173 (2019)

\bibitem {Buda}H.A. Buchdahl, Mon. Not. Roy. Astron. Soc. 150, 1 (1970)

\bibitem {bd1}C. Brans and R.H. Dicke, Phys.\ Rev. 124, 925 (1961)

\bibitem {cl1}T.\ Clifton and J.D. Barrow, Phys. Rev D 72, 103005 (2005)
[Erratum: Phys. Rev. D 90, 029902 (2014)]

\bibitem {cl2}A. De Felice and S. Tsujikawa, Living Reviews Relati. 13, 3 (2012)

\bibitem {cl3}I. Roxburgh, Gen. Rel. Grav. 8, 219 (1977)

\bibitem {anbas1}A. Paliathanasis, M.\ Tsamparlis and S. Basilakos,
Phys.\ Rev. D 84, 123514 (2011)

\bibitem {sd3}J.D. Barrow and A.C. Ottewill, J. Phys. A 16, 2757 (1983)

\bibitem {gl1}G. Leon and E.N. Saridakis, Class. Quantum Grav. 28, 065008 (2011)

\bibitem {gl2}G. Leon, Int. J. Mod. Phys. E 20, 19 (2011)

\bibitem {Ferraro}R. Ferraro and F. Fiorini, Phys. Rev. D 75, 084031 (2007)

\bibitem {sd4}S. Basilakos, S. Capozziello, M. De
Laurentis,\ A.\ Paliathanasis and M.\ Tsamparlis, Phys.\ Rev. D 88, 103526 (2013)

\bibitem {sd2}A. Paliathanasis, Mod. Phys. Lett. A 36, 2150261 (2021)

\bibitem {f6}J. B. Jimenez, L. Heisenberg and T. Koivisto, Phys. Rev. D 98,
044048 (2018)

\bibitem {ff1}L. Atayde and N. Frusciante, Phys. Rev. D 104, 064052 (2021)

\bibitem {ff2}R. Solanki, A. De and P.K. Sahoo, Phys. Dark Universe 36, 100996 (2022)

\bibitem {ff3}A. Lymperis, JCAP 11, 018 (2022)

\bibitem {ff4}S.H. Shekh, Phys. Dark Univ. 33, 100850 (2021)

\bibitem {ff4a}A. Kumar Singha, A. Sardak and U. Debnath, Phys. Dark. Universe
41, 101240 (2023)

\bibitem {ff4b}S.A. Narawade, Laxmipriya Pati, B. Mishra, S.K. Tripathy, Phys.
Dark. Universe 36, 101020 (2022)

\bibitem {ff5}M. Koussour, S. Arora, D.J. Gogoi, M. Bennai and P.K.\ Sahoo,
Nucl. Phys. B 990, 116158 (2023)

\bibitem {ff6}T.-H. Loo, M. Koussour and A. De, Annals Phys. 454, 169333 (2023)

\bibitem {ff7}M. Koussour, S. Dahmani, M. Bennai and T. Ouali, Eur. Phys. J. C
138, 179 (2023)

\bibitem {nn5}K. Hu, T. Katsuragawa and T. Qiu, Phys. Rev. D 106, 044025 (2022)

\bibitem {nn7}M Koussour, S.K.J. Pacif, M. Bennai and P.K. Sahoo, Fortschritte
der Physik 71, 2200172 (2023)

\bibitem {ww2}W. Wang, H. Chen and T. Katsuragawa, Phys.\ Rev. D 105, 024060 (2022)

\bibitem {qq2}N. Dimakis, A. Paliathanasis and T. Christodoulakis, Class.
Quantum\ Grav. 38, 225003 (2021)

\bibitem {qq3}N. Dimakis, M. Roumeliotis, A. Paliathanasis, P.S.
Apostolopoulos and T. Christodoulakis, Phys.\ Rev. D 106, 123516 (2022)

\bibitem {ffq1}A.S. Agrawal, B. Mishra and P.K. Agrawal, Eur. Phys. J. C 83,
113 (2023)

\bibitem {ffq2}S. K. Maurya, Ksh. Newton Singh, Santosh V Lohakare and B.
Mishra, Fortschr. Phys. 70, 2200061 (2022)

\bibitem {ffq3}S.A. Narawade and B.Mishra, Annalen der Physik, 535, 2200626 (2023)

\bibitem {ffq4}S.A. Narawade, S. H. Shekh, B. Mishra, W. Khyllep and J. Dutta,
Constraining parameters for the accelerating models in symmetric teleparallel
gravity, (2023) [arXiv:2303.01985]

\bibitem {ffq5}S.A. Narawade, Shashank P. Singh and B. Mishra, Physics of the
Dark Universe 42, 101282 (2023)

\bibitem {qq4}A. Paliathanasis, Phys. Dark Univ. 41, 101255 (2023)

\bibitem {qq4a}W. Khyllep, A. Paliathanasis and J. Dutta, Phys. Rev. D\ 103,
103521 (2021)

\bibitem {qq5}C. Bohmer, E. Jensko and R. Lazkoz, Universe 9, 166 (2023)

\bibitem {qq6}H. Shabani, A.\ De and T.-H. Loo, Phase-space analysis of a
novel cosmological model in f(Q) theory, (2023) [arXiv:2304.02949]

\bibitem {ds00}J. Shi, Cosmological constraints in covariant f(Q) gravity with
different connections, (2023) [arXiv:2370.08103]

\bibitem {rev10}L. Heisenberg, Review on f(Q) Gravity, (2023) [arXiv:2309.15958]

\bibitem {Heis2}F. D' Ambrosio, L. Heisenberg and S. Kuhn, Class. Quantum
Grav. 39 025013 (2022)

\bibitem {ds1}D. Zhao, Eur. Phys. J. C 82, 303 (2022)

\bibitem {ref1}F. Bajardi and S. Capozziello, Minisuperspace quantum cosmology
in f(Q) gravity, (2023) [arXiv:2305.00318]

\bibitem {cop1}E. J. Copeland, M. Sami and S. Tsujikawa, Int. J. of Mod. Phys.
D 15, 1753,(2006)

\bibitem {cop2}L. Amendola and S. Tsujikawa, Dark Energy Theory and
Observations, Cambridge University Press, Cambridge UK, (2010)

\bibitem {cop3}A. Paliathanasis, M. Tsampalis, S. Basilakos and J.D. Barrow,
Phys. Rev. D 91, 123535 (2015)

\bibitem {cop4}C. Rubano and J. D. Barrow, Phys. Rev. D. 64, 127301 (2001)

\bibitem {cp5}V. Sahni and A. Starobinsky, Int. J. Mod. Phys. D 9, 373 (2000)

\bibitem {arx1}H. Shabani, A. De, T.-H. Loo and E.N. Saridakis, Cosmology of
f(Q) gravity in non-flat Universe, (2023) [arXiv:2306.13324]

\bibitem {per1}J.B. Jimenez, L. Heisenberg, T.S. Koivisto and S. Pekar, Phys.
Rev. D 101, 103507 (2020)

\bibitem {per2}A. Najera and A. Fajardo, JCAP 03, 020 (2022)
\end{thebibliography}
\end{document}